\documentclass[10pt,conference,final]{IEEEtran}

\usepackage[utf8]{inputenc}
\usepackage[T1]{fontenc}

\usepackage{clrscode}
\usepackage{xcolor}
\usepackage[pdftex]{graphicx}

\usepackage[hyphens]{url} %

\definecolor{DarkBlue}{rgb}{0.1,0.1,0.7}

\definecolor{DarkGreen}{rgb}{0.1,0.5,0.1}

\newcommand{\documenttitle}{The User Behavior in Facebook and its Development from 2009 until 2014}

\definecolor{thomas}{RGB}{0,100,0}

\definecolor{thorsten}{RGB}{0,0,255}

\definecolor{daniel}{RGB}{200,10,0}

\begin{document}

\title{\documenttitle}

\author{
  \IEEEauthorblockN{Thomas Paul$^+$, ~ Daniel Puscher$^+$, ~ Thorsten Strufe$^*$}
  \IEEEauthorblockA{TU Darmstadt$^+$ ~ and ~ TU Dresden$^*$\\
thomas.paul@cs.tu-darmstadt.de, uni@daniel-puscher.de, thorsten.strufe@tu-dresden.de
   } 
}

\maketitle

\begin{abstract}

Online Social Networking is a fascinating phenomena, attracting more than one billion people. It supports basic human needs such as communication, socializing with others and reputation building. Thus, an in-depth understanding of user behavior in Online Social Networks (OSNs) can provide major insights into human behavior, and impacts design choices of social platforms and applications. However, researchers have only limited access to behavioral data. As a consequence of this limitation, user behavior in OSNs as well as its development in recent years are still not deeply understood.

In this paper, we present a study about user behavior on the most popular OSN, Facebook, with 2071 participants from 46 countries. We elaborate how Facebookers orchestrate the offered functions to achieve individual benefit in 2014 and evaluate user activity changes from 2009 till 2014 to understand the development of user behavior. Inter alia, we focus on the most important functionality, the newsfeed, to understand content sharing amongst users. We (i) yield a better understanding on content sharing and consumption and (ii) refine behavioral assumptions in the literature to improve the performance of alternative social platforms. Furthermore, we (iii) contribute evidence to the discussion of Facebook to be an aging network.%

\end{abstract}


\maketitle

\section{Introduction}

Online Social Networks (OSNs) offer a rich set of different communication and content sharing functionalities to their users and change the way how people interact. They support basic human needs such as communication, socializing with others and reputation building. Facebook\footnote{https://facebook.com/; Accessed 2015-05-06} is the dominating social networking service in the world that attracts more than one billion people\footnote{http://allfacebook.de/userdata/; Accessed 2015-05-06}. Because of their importance for both, the individual users and the modern society, OSNs are subject in many research areas. Inter alia, researchers examine %
user interactions \cite{Jiang}, the information diffusion \cite{bakshy2012role} or novel social networking architectures e.g.  \cite{Cutillo3,Buchegger,my3}. An in-depth understanding of user behavior in OSNs can provide major insights into human behavior, and impacts design choices in social platforms and applications. However, user behavior in OSNs is still not deeply understood, because researchers have only limited access to behavioral data.

To this end, we present a study to understand how users orchestrate Facebook's functions to gain benefit for themselves as well as to understand the benefit that the most important function (Figure \ref{fig:functionalities}), the  newsfeed, provides to Facebook users. We also take the changes of user behavior from 2009 till 2014 into account to understand the success and aging process of Facebook, and compare our findings with user behavior assumptions in the literature.

Our study is based on data, which is collected at the client-side. We gathered it from 2,071 users via web-browser plug-in to overcome limitations of crawled datasets (\cite{Catanese:crawler,Meo:2014:AUB:2542182.2535526,jiang2013understanding,Gyarmati}), click streams \cite{Schneider} or social network aggregator data \cite{Benvenuto}. Our plug-in is able to measure client-side activity such as scrolling or deactivating tabs to estimate the time that users invest to examine newsfeed posts. The plug-in has access to profile details endowed with user's rights and is able to read activity logs that encompass historical actions regardless of their origin from mobile or stationary devices.  

To find volunteers for this study who are eager to install our plug-in, we sent a solicitation to join this study to users of our previous work \cite{paul2012c4ps} where we created a new interface to simplify audience selection. This interface is based on a color coding and minimizes the effort as well as error rates when changing the visibility of content items. We published a browser extension (plug-in) for Firefox and Chrome, called Facebook Privacy Watcher (FPW), which implements this new type of interface on purpose to yield benefit to people. Both versions (Chrome and Firefox) of the FPW have together been installed by more than 44,000 Facebook users.

We asked the FPW users to join a user study about user behavior in Facebook by installing a second browser extension that anonymously collects the data for the study which is presented in this paper. 2,071 FPW users allowed us to evaluate their user behavior in detail. We collected basic demographical data such as gender and age, data on usage patterns with respect to functionalities, data about communication partners w.r.t. to the social graph distance as well as metadata about the content which is shared. This metadata consists of the type of content, the time when it has been created or watched as well as its size in bytes (if available). We respect the privacy of all probands by not storing or evaluating any content or identifier!

To understand the user behavior on Facebook, we evaluate the dataset with focus on the questions:  How do people orchestrate the vast variety of functions? Who produces content in Facebook? How much and which kind of information do people share on Facebook? Who consumes which content? How old is the shared content until it is viewed and how long is it commonly consumed? How does the observed user behavior change over time?

The main findings are that Facebook sessions are very short, compared with assumptions in the literature %
and users' content contributions are extremely disparate in type and quantity. A major share of newsfeed stories is posted by a minority of users and consists of reshared, liked or commented issues rather than original user-generated content. Facebook manages to compensate this lack of high quality content by transforming commercial posts into regular newsfeed content that is accepted by FPA users equally beside user-generated content. Evaluating the history of user actions from 2009 till 2014, we show that users tend to befriend with less other users. Also, actions that cause little effort, such as reshares and likes, recently became more popular than status updates and comments.

With this work, we contribute to better understand the usage of Facebook with respect to churn, content contribution and consumption, as well as communication patterns. We also help authors of alternative (e.g. P2P-based) OSN architectures to make well-founded design choices and provide evidence for Facebook to be an aging network by analyzing dynamics of user behavior over time.

The remainder of the paper is structured as follows: We discuss related works in Section \ref{sec:related_work} and describe the experimental setup in Section \ref{sec:experimental_setup}. We examine the attention that users pay to Facebook by evaluating the session durations and frequencies (churn) in Section \ref{sec:churn}, and evaluate the popularity of different functionalities in Facebook in Section \ref{sec:function_popularity}. We elaborate the newsfeed with respect to content creation, composition and consumption in Section \ref{sec:news_feed}, evaluate communication patterns of FPA users in Section \ref{sec:communication_patterns} and examine dynamics in user behavior in Section \ref{sec:dynamics}. In Section \ref{sec:conclusion}, we summarize our work and draw major conclusions.

\section{Related Work} 
\label{sec:related_work}

A vast amount of related work in the field of user behavior in OSNs has been done by now. Prior work can be classified by the research discipline (e.g. computer science and psychology), the data collection method as well as the social network that has been analyzed (e.g. Twitter, Facebook, Google+).

We aim to understand what is the technical core of Facebook's success. In the remainder of this section, we thus focus on related work in computer science about Facebook that evaluates functionality rather than e.g. the social graph. Unlike works in the field of psychology or sociology, we do not aim to contribute in exploring OSN user's individual properties or any relation between OSNs and societies. We take the system's point of view and want to know how users use Facebook's functionality and how to build excellent systems that support user's needs. Since capturing functionality usage causes stringent requirements on the data collection process, we examined related work based on this criteria in the remainder of this Section.

Many works on user behavior are based on crawler-gathered data \cite{Catanese:crawler,Meo:2014:AUB:2542182.2535526,jiang2013understanding,Gyarmati}. Datasets which are acquired by employing crawlers contain static elements of user profiles. Dynamics can be estimated by frequently crawling the same information to detect changes. Also activity counters such as page view counters can detect some types of dynamics \cite{strufe2010profile}. 
Page view counters have been also used by Lin et al. \cite{lin2012analysis}. They crawled Renren and Sina which offer pageview counters and allow to crawl a list of the last nine visitors. This type of information allows insights into profile visits by building directed and weighted graphs. However, %
datasets which are acquired by employing crawlers are not sufficient for our purpose to understand how users use Facebook since they neither reflect the use of all kinds of functionality (such as messaging, likes or the 'Timeline' usage) or their interplay nor do they allow to evaluate exact timings.

Schneider et al. \cite{Schneider} analyzed passively monitored click streams of Facebook, MySpace, LinkedIn, Hi5, and StudiVZ. They analyzed feature and service popularity (churn), click sequences and profile usage but do not evaluate information about the newsfeed and profile page compositions or historical data.  We can confirm the finding that users are trapped when using a specific functionality. If users view pictures, the likelihood is extremely high that the next action is again to view pictures. However, Schneider et al. did not evaluate content consumption nor content contribution habits of users.

Benvenuto et al. \cite{Benvenuto} %
elaborated click stream data from 37,024 users of Orkut, MySpace, Hi5, and LinkedIn in a twelve days period, collected by a Brazil social network aggregator. They additionally analyzed crawler data which was collected at Orkut. The main results of this work are session descriptions containing information such as how long and how often users use which function of the system. With the help of the crawler data on Orkut, Benvenuto et al. analyzed function usage with respect to friend relations. In contrast to this work, we focus on Facebook in 2014. Due to client-side data collection, we know timing issues such as pre-click times, can distinguish between stranger profiles and professional sites and are able to do long-term evaluations based on activity log data. Another aggregator study in the middle east that measures session properties such as session durations can be found in \cite{shahrak2014middle}.

To overcome these dataset implicated downsides in understanding Facebook usage, Facebook-internal applications and client-side data collection methods have been developed. Mondal et al. \cite{mondal2014understanding} leveraged a Facebook app to examine Social Access Control Lists and Luarn et al. \cite{Luarn20141} developed a Facebook app to test and confirm the hypothesis that people's network degree is positively correlated with the the frequency of information dissemination. Client-side data collection can be found at \cite{weinreich2006off,velayathan2007investigating}. However, these works do not analyze OSNs but web surfing behavior in general. 

The allocation of attention amongst friends has been analyzed by Facebook \cite{backstrom2011center}. The main findings are that Facebook users concentrate their attention on a small fraction of friends while messaging is much more focused on few individuals than profile page views. Backstrom et al. observed a gender homophily: ``We find that females send 68\% of their messages to females, while males send only 53\% of their messages to females. This distinction is consistent with gender homophily — in which each gender has a bias toward within-gender communication — modulated by the overall distribution of Facebook messages. On the other hand, we see much smaller differences in viewing: for typical activity levels, both females and males direct roughly 60\% of their profile viewing activity to female users.'' 
In contrast to Backstrom et al. \cite{backstrom2011center}, we %
evaluate user behavior patterns more detailed with respect to timings and content contribution and consumption. We further evaluate historical and device usage independent data to understand the development of user behavior over time.%

\section{Experimental Setup}
\label{sec:experimental_setup}

In this Section, we describe our ethical considerations regarding this study, the data collection process, the amount and the composition of the data that we collected. We further describe the bias of the data with respect to the differences to the complete set of Facebook users.

\subsection{Ethical Considerations}

We acquired our participants by asking FPW users to participate in this study. Before installing the FPA, we explained the reason for collecting the data to our study participants and allowed users to access and verify all data before sending it to our server with consent. We further did not violate any rule on Facebook since we directly gathered our data from our participant's browser.

All data that we used for this study is anonymized and encrypted for transmission with state-of-the-art technology. We did not collect or store any content or messages but metadata about the user behavior such as content types, time stamps and hashes of ids to be able to distinguish amongst actors without being able to identify individuals. Nevertheless, we keep the collected data confidential to protect all study participants from deanonymization attempts and do only publish aggregated data.

\subsection{Sample Generation}

We published the browser extension (FPW) to simplify audience selection for posts on Facebook. %
Enclosed into an FPW update, we asked our users whether they would be willing to help us with a user behavior study and thus share the necessary data with us. 11,572 FPW users filled out the questionnaire: 11.8\% of the users answered ``yes'' 21.3\% of the user ``maybe'' and the rest answered with ``no''. This questionnaire was intended to be a risk reducing pre-test before developing the data collection plug-in to make sure that a reasonable number users are willing to help us.

Subsequently to this successful pretest, the development of the data collection plug-in started as a statistic module of the Facebook Privacy Watcher and was thus called Facebook Privacy Analyzer. However, we decided the FPA to be a stand-alone plug-in.%

The FPA version for the Chrome browser was published at Chrome web store on 24th of November 2013. After fixing several bugs, the Firefox Version was published on 16th of January 2014. FPW users who have previously agreed to join the study have immediately been asked to install the plug-in by showing a pop-up window in FPW. All other FPW users were asked a second time again two weeks later. To incentivise study participation, the data collecting plug-in FPA provided statistics to users about their own behavior. The statistics also could be shared on Facebook and compared with friends.

\subsection{Sample Bias}

Since we cannot force any randomly chosen person to join our study, we only studied persons which were eager to help us in doing research. The data that we collected is thus neither a result of a random sampling process nor the complete Facebook dataset. We thus suffer from a bias that is evaluated in this Section.

The majority of our participants joined us by following an invitation via pop-up message in the FPW. The FPW became popular by newspaper articles and radio station broadcasts that reached even less technology-savvy people of all ages. Beside some international media such as ``golem'' or ``Der Standard'', the center of FPW news coverage was in Germany.
Hence, the overwhelming majority of the FPA users are originated from Germany (84.87\%), too. The rest of the participants are - according to the information in their profiles - from 45 other countries. 

\begin{figure}[ht]
\centering
\includegraphics[width=0.4\textwidth]{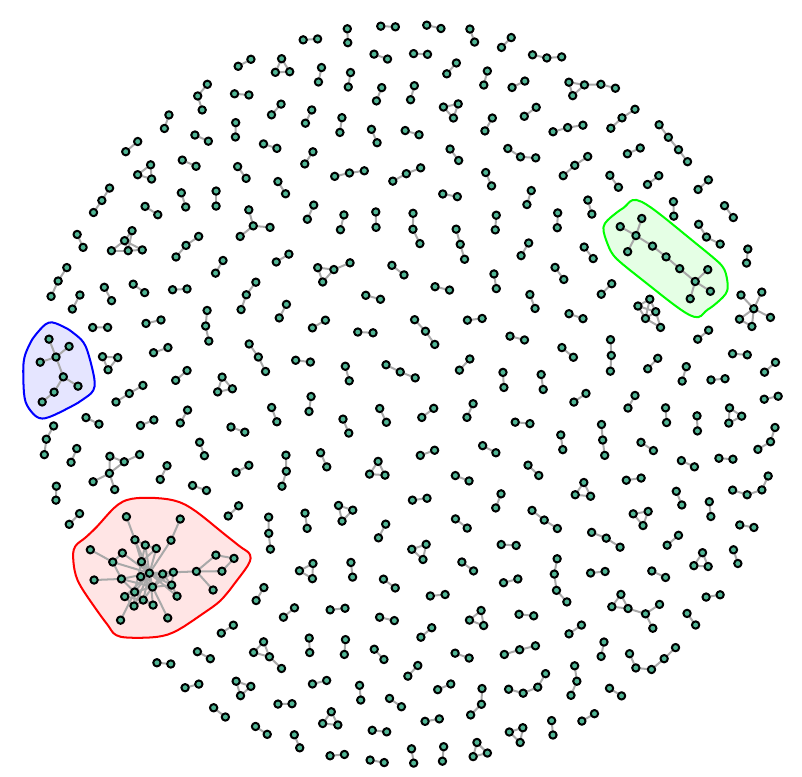}
\caption{Friend relations amongst FPA users; singletons are excluded}
\label{fig:friendgraph}
\end{figure}

The heterogeneity of our sample is illustrated by Figure \ref{fig:friendgraph}. It shows the friendship relations amongst 621 FPA users. Study participants without friends who also use the FPA (singletons; 68.02\%) are not part of this graph. Only 3 bigger clusters of 31, 11 and 8 users exist. We assume the biggest clusters to consist of persons acquainted with one of the authors.

To further estimate the sample bias, we recorded the gender and the year of birth of our participants from their user profiles. 77.6\% of the participants are male, 21.13\% are female and the rest did not share this information with us. Some users claimed to be born before 1925 or after 2010. However, we assume the majority of age information to be correct. The majority of our participants are born between 1960 and 1985. Our median participant is a 44 years old male Facebook user. Compared with average Facebook users, our participants are older and males are overrepresented.

\subsection{Details on Collected Data}

To understand the Facebook usage, we require data that describes both components of interaction together with the respective timing information: the actions performed by users as well as the information flow from Facebook to the users. We mainly collected four data types, using the FPA: the performed actions of users, the friend lists of users, activity logs and basic demographic information about the FPA users. Furthermore, we measured exact session durations by storing the time when activating and deactivating browser tabs in case of active Facebook sessions. In the remainder of this section, we explain the four main types of collected information and their necessity for our study.

\paragraph{Performed Actions}
Every action that a user performed in Facebook has been recorded in the database, together with a timestamp and the browser tab ID. We recorded further metadata about the actions such as the hashes of persons who are involved in those actions. 

\paragraph{Friend Lists}
We recorded the friend list as set of UIN hashes. We needed this information for several reasons: We checked whether two-sided actions such as messaging or profile views are performed amongst friends or strangers, we counted the number of friends of each FPA user to calculate the node degree within the ego-graph and we checked newsfeed posts whether they are originated from friends.

\paragraph{Activity Logs}
\label{par:activity_log}
The FPA is a powerful tool. However, it can only gather data in case the user uses a web browser based Facebook access. Facebook maintains an activity log as a part of the user profiles. This activity log contains activity records back till the time of registration at Facebook independent from the access channel (e.g. mobile app or browser). 

These records contain almost all actions which have been performed on Facebook together with the timestamp and some metadata such as communication partners. Private messaging e.g. is not included in the activity logs. Nevertheless, it is a very valuable data source for our analysis since it allows us to estimate the fraction of actions that we can observe in the browser. We can thus bridge the gap that would appear in case of only evaluating data from Firefox or Chrome browsers. 

Furthermore, due to the activity log's long term records, we can evaluate changes in user behavior during time. We can thus trace the learning process of new users joining Facebook.

\paragraph{User Demographics}

Based on ethical considerations to protect user's privacy, we only stored basic user data such as age and gender. We need this data to estimate the bias of our sample.

\subsection{Data Quantification}

Since we changed parts of the code basis through updates after the first publication of the plug-in, we decided to use only data that has been collected after 1st of January 2014. During our observation period of 123 days, 2071 users installed the FPA. However, not every study participant joined at the first day and not every participant stayed the whole rest of the time. We thus observed our participants on average 34 days. %

\subsection{Mobile Device Usage}
As explained above, the activity logs contain even those actions which have been performed on mobiles devices or stationary devices without an installed FPA instance. Furthermore, the information flow from Facebook to users (e.g. the newsfeed) is independent from the devices that have been used to create and post the original content. We thus can include mobile actions in the majority of our analysis. However, we can neither evaluate content consumption nor churn on mobile devices.  

\section{Churn}
\label{sec:churn}

Churn denotes in this work the behavior of users starting and ending Facebook sessions. In this Section, we describe the churn behavior that we observed to provide insights into the importance of Facebook for the FPA user's daily life.

Caused by the properties of web-based systems in which communication is triggered by user activity (events), different methods exist to measure churn. The related work (e.g. \cite{Schneider}) use inter alia the absence of activity (timeouts) as an indicator for users to leave the system. In contrast to the related work, the browser plug-in FPA has access to more precise information such as whether a tab is activated or not. To ensure comparability, we include four different measurements instead of only presenting the most precise measurements. We distinguish amongst the following four measurement methods:

\begin{itemize}
 \item \textbf{Basic: } A session always starts with the login on Facebook and ends with either a logout or with closing the last open browser tab with an open Facebook session. However, this session definition leads to extreme cases of sessions lengths lasting several days. It does not realistically reflect the user's attention. We thus included a timeout of three hours starting after the last action was performed by the user.
 \item \textbf{Timeout: } The timeout measurements are conform to the previous measurements, using a more aggressive timeout of 5 minutes. We ague that this short timeout reflects user attention better than previous. 
 \item \textbf{Basic without timeout: } We included the previous measurements without timeout to quantify the effect of the timeouts on our basic measurements. Comparing 'basic' and 'basic without timeout' indicates how many users log themselves off or close Facebook tabs while leaving.
 \item \textbf{Precise: } Since the FPA notices tabs to be activated and deactivated, the most precise measurement is to count a session to start as soon as either a users performs a login action on Facebook or a browser tab on Facebook is activated. The session ends in case the browser (or the tab) is either closed or deactivated or a logout action is performed. 
\end{itemize}

\begin{figure}[ht]
\centering
\includegraphics[width=0.489\textwidth]{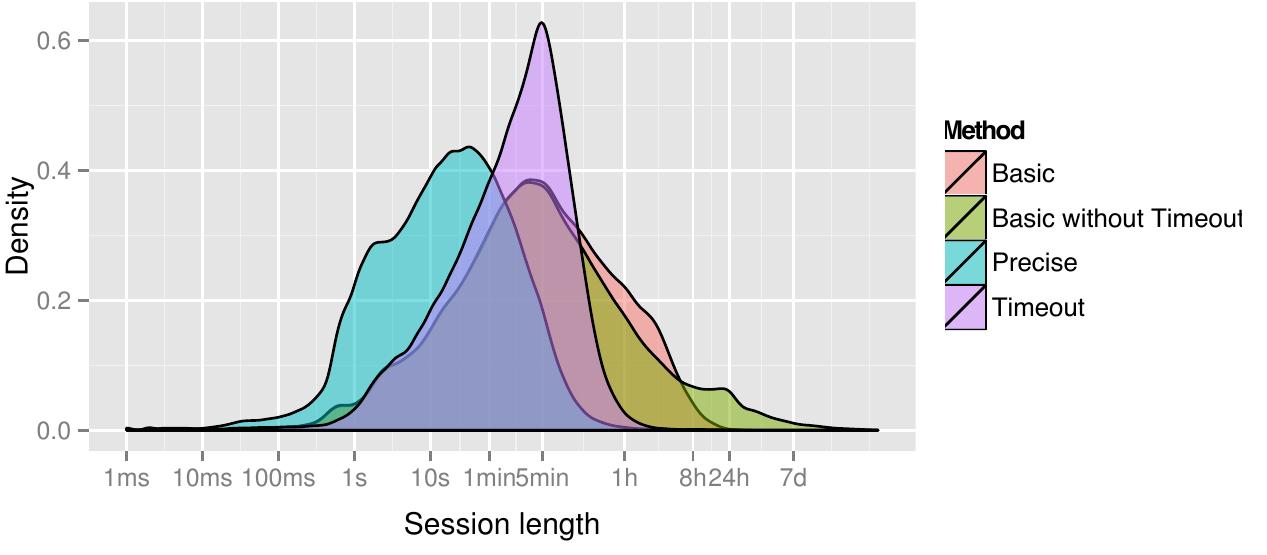}
\caption{Distribution of session durations with respect to four different measurement methods on a logarithmic scale}
\label{fig:session_length_probability_density}
\end{figure} 

Figure \ref{fig:session_length_probability_density} shows the duration of sessions. The 'precise' and 'timeout' measurements depict very short Facebook sessions of 2:16 minutes on average (median: 0:17) for 'precise' and an average of 5:32 (median 2:21) for 'timeout'. The  average results of the 'basic' measurements (31:40 minutes) are roughly in line with the results of Schneider et al. \cite{Schneider} and the measurements without timeout show unrealistic lengths of 240:01 minutes.

\begin{figure}[ht]
\centering
\includegraphics[width=0.489\textwidth]{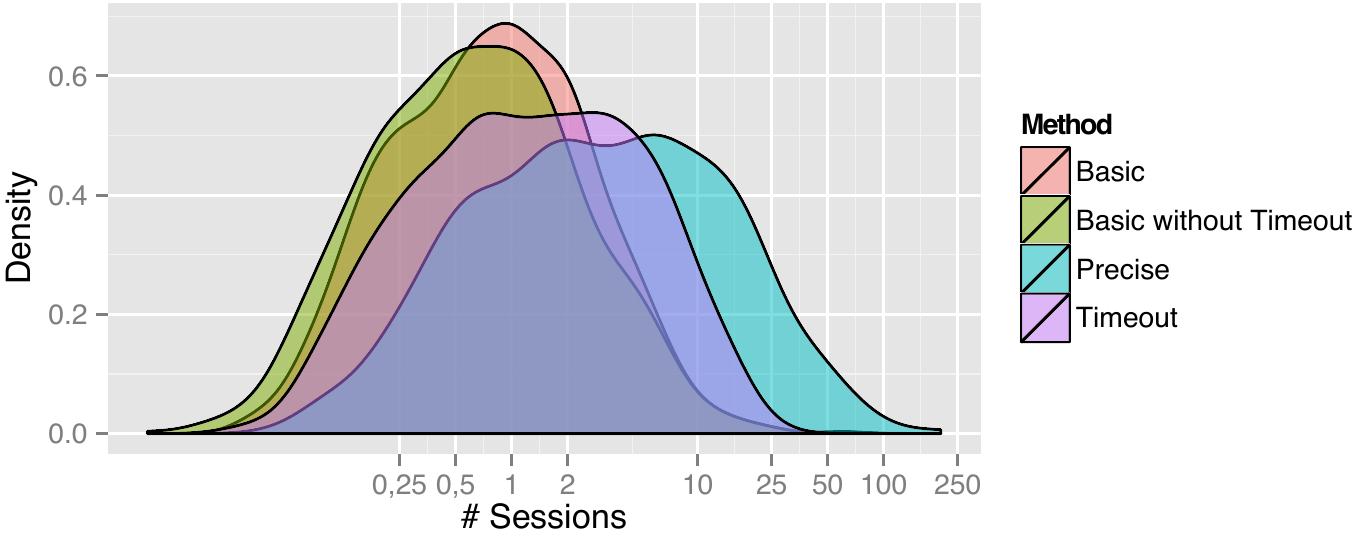}
\caption{Average number of sessions per day}
\label{fig:session_average _number}
\end{figure} 

Beside the session duration, the average number of sessions is important, too. Figure \ref{fig:session_average _number} shows the distribution of average session numbers. The 'precise' measurement indicates an average of 2.97 sessions, the 'timeout' method 1.28 and the 'basic' indicates an average of 0.79 sessions per day.

The observed session duration as well as the total online duration per day (average session duration $\times$ average number of sessions) are shorter than those in the evaluation assumptions of many Peer-to-Peer based decentralized OSNs (e.g. \cite{my3,soup,s-data}). Our dataset provides evidence that assuming shorter session durations, the absence of stable nodes as well as an average of less than three sessions per day may be more realistic.

\section{Function Popularity}
\label{sec:function_popularity}

\begin{figure}[b]
\centering
\includegraphics[width=0.489\textwidth]{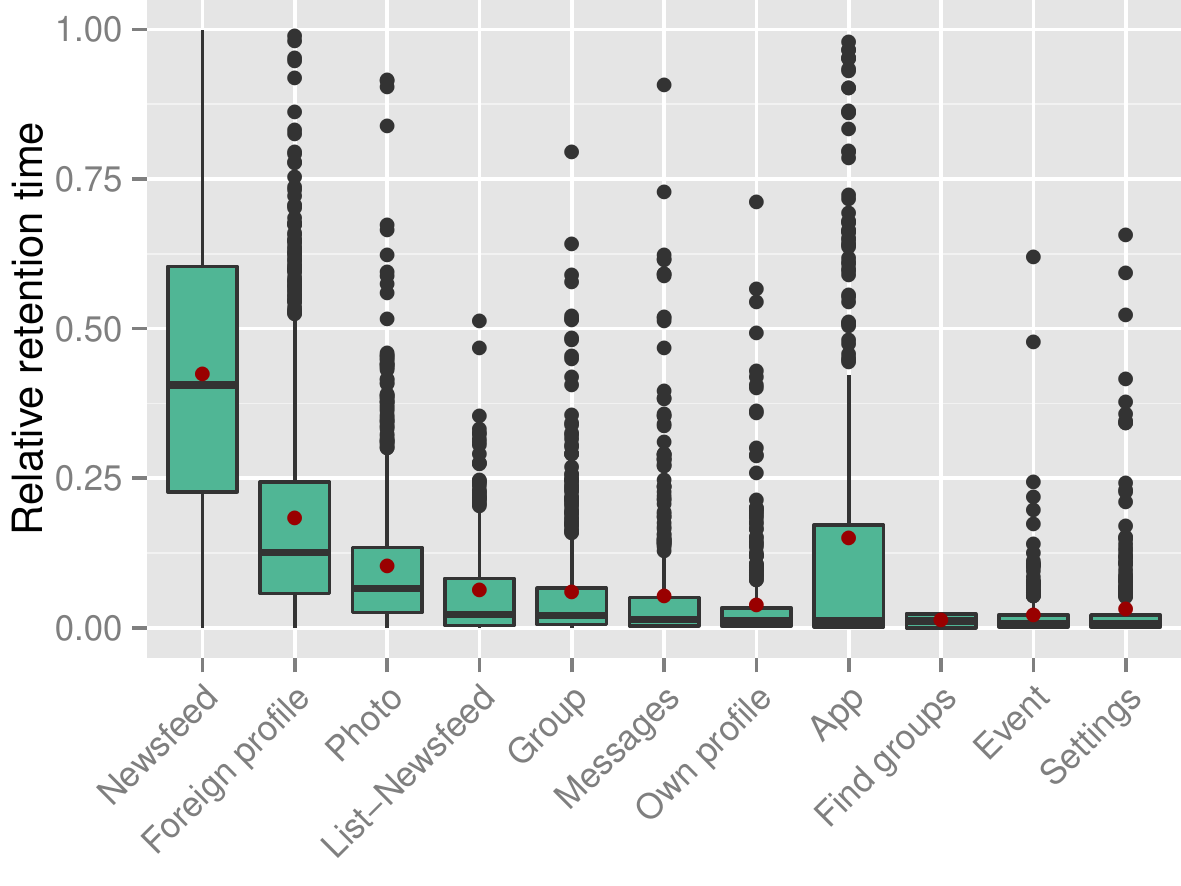}
\caption{Fraction of time that FPA users spent using different Facebook functionalities}
\label{fig:functionalities}
\end{figure}

Beside its third-party app ecosystem, Facebook itself comprises a rich compilation of functions which characterize the service. Figure \ref{fig:functionalities} contains box-whisker plots which show the relative fraction of time that users spend with each function. The box-whisker plots are ordered by the median time that users spend with each function. 

The newsfeed, called 'Timeline', dominates Facebook usage, followed by viewing other user's profiles. Viewing pictures is very popular, too. Surprisingly, users spend more time with topic-related interest groups (named 'groups' in Figure \ref{fig:functionalities}) and the enclosed newsfeeds ('list newsfeed') than with exchanging messages with others. The median FPA user spends more time with maintaining the own profile than with running apps on Facebook. However, running apps is not unpopular in general. A minority of users spends a big fraction of their time with apps. 

\begin{figure}[ht!]
\centering
\includegraphics[width=0.489\textwidth]{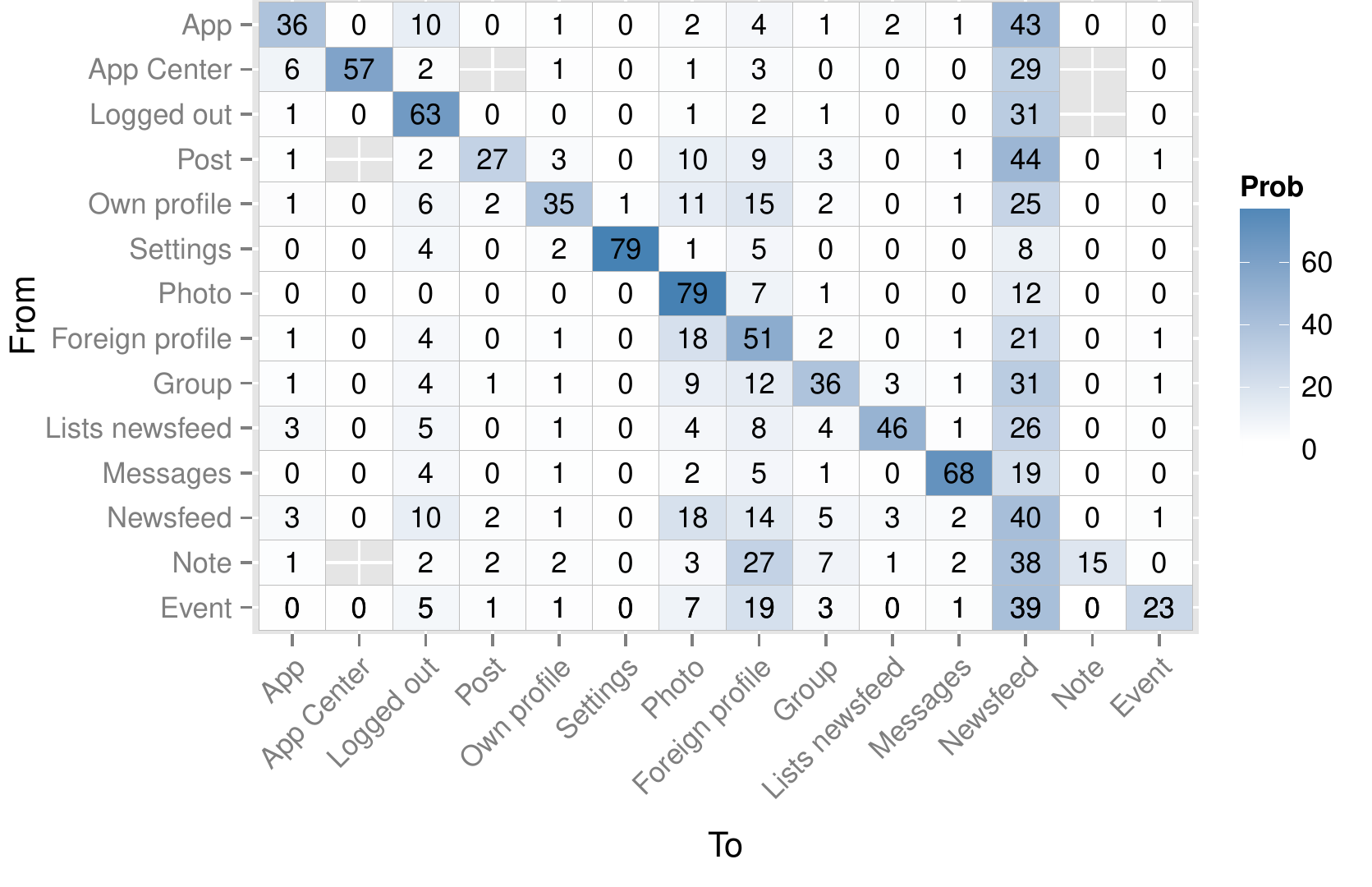}
\caption{Page transition matrix; the row values sum up to 100 \%}
\label{fig:page_transitions_matrix}
\end{figure}

The page transition matrix in Figure \ref{fig:page_transitions_matrix} shows relations between different functions in Facebook by depicting page type transitions. Two main findings can be quickly realized: Users tend to repeat actions several times (e.g. view a picture after viewing a picture) and the newsfeed is the dominating functionality getting the most page hits from other transition sources.

\subsection{Third Party Applications}

In this section, we evaluate the usage of third party applications (apps) which are completely integrated into Facebook itself and leverage the Facebook platform to provide benefit to their users. For these evaluations, we only included data from users who never accessed their newsfeed to exclude strangers and those who joined our experiment for less than one week. We thus only used the data of 1,068 users.

The mutual benefit of the app creator and Facebook are that Facebook's functionality is extended by third parties and the third parties can leverage Facebook platform functionality. Based on the platform functionality, the App instances of different users can communicate with each other and the platform allows the apps to receive information about the users, such as the friendship connections or interests, in case of user's consent. The apps can thus leverage the social graph to fortify collaboration amongst friends and to establish or support a feeling of togetherness.  

\begin{figure}[ht]
\centering
\includegraphics[width=0.489\textwidth]{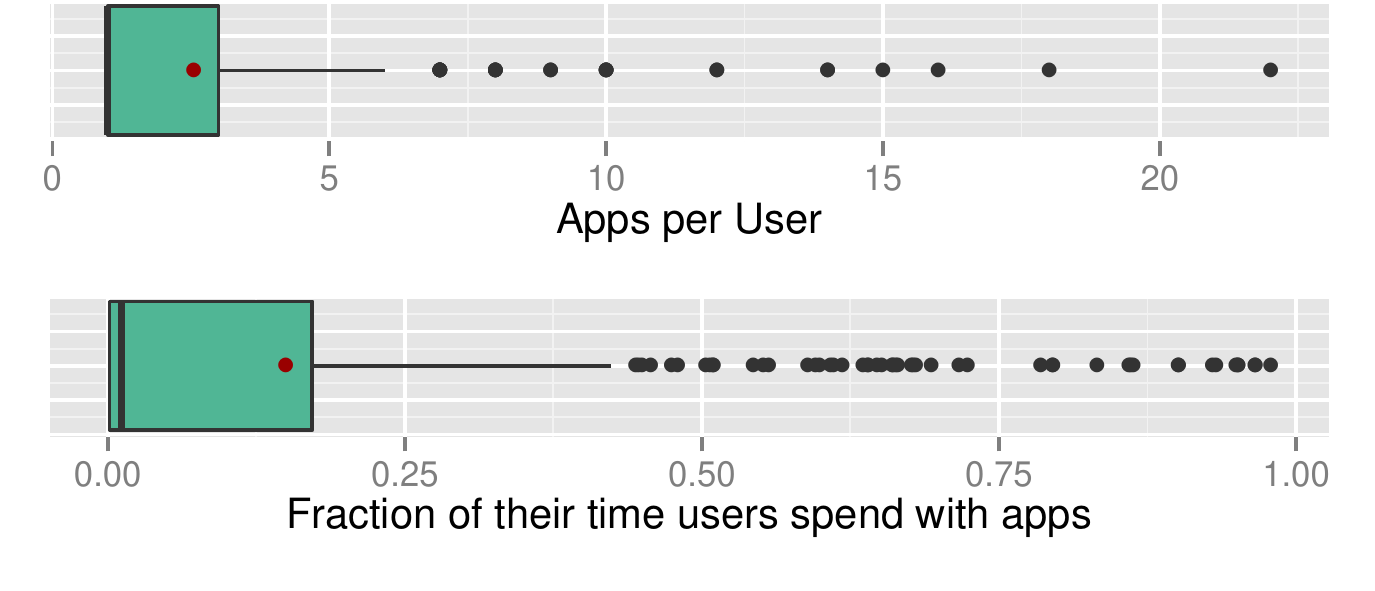}
\caption{Box-whisker-plots: distributions of the number of apps FPA users are using and the time they are spending with apps}
\label{fig:app_stats}
\end{figure}

29.96\% of the FPA users used at least one app and 50,88 \% of all app users used exactly one app. On average, app users use a total number of 2.55 apps and spend 14.97 \% of their time on Facebook with running apps. Figure \ref{fig:app_stats} shows the distributions of the number of apps the users from the evaluation set are using as well as the distribution of the fraction of time they spend. 

\begin{figure}[ht]
\centering
\includegraphics[width=0.4\textwidth]{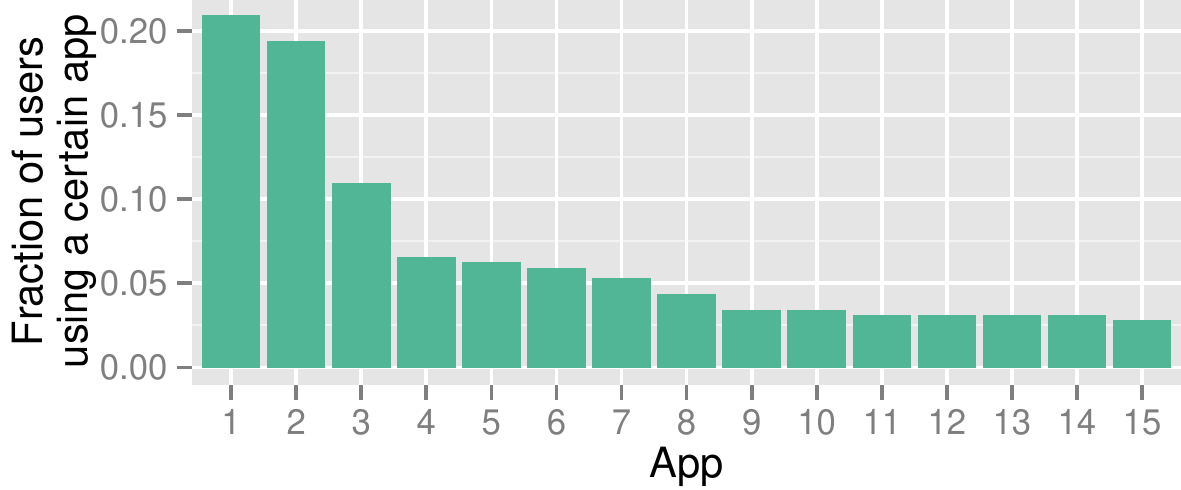}
\caption{Popularity of apps; caused by privacy limitations, we do not know the names of the apps and thus numbered them}
\label{fig:app_popularity}
\end{figure}

Unfortunately, because of our privacy limitations, we recorded the hash values of app IDs rather than their name. We hence cannot evaluate which application is the most popular one. In spite of this limitation, Figure \ref{fig:app_popularity} shows the popularity distribution of apps amongst users. We can learn from that figure that the overwhelming portion of app usage concentrates on very few apps. The most popular two applications each are used by 20\% of all FPA users and the 3rd popular application is only use by roughly half that many users.

Evaluating the app usage suggests that Facebook is not mainly a gaming platform. This result is in line with our session duration observations in Section \ref{sec:churn}. However, a few third-party applications are used by roughly 30\% of the users.

\section{The Facebook Newsfeed}
\label{sec:news_feed}

Being the core of Facebook, we dedicate this Section to the Timeline. Attracting users to generate and contribute content to the Timeline, such as pictures, videos status updates and text messages, is crucial for the success of social networking platforms like Facebook. However, the platform operator needs to solve a chicken-and-egg problem: content contributors need to be incentivised to contribute content by the existence of an audience that is interested in their submissions and the crucial incentive for a potential audience to spend their attention to the platform is the existence of interesting content. 

Thus, Facebook's role is to be an information mediator that manages two scarce resources: valuable and interesting content as well as attention of the audience. ``The goal of News Feed is to deliver the right content to the right people at the right time so they don’t miss the stories that are important to them. Ideally, we want News Feed to show all the posts people want to see in the order they want to read them.`` \cite{window_timeline}

The straightforward way to only leverage friendship connections as communication channels while respecting restrictions arising from privacy settings is not sufficient to find an interested audience for content. The matchmaking is a hard task to solve for two reasons: a minority of users posts a lot of content which is not interesting for all of their friends and the amount of posts quickly becomes too high to be read by others \cite{window_timeline}. %
Facebook thus decides which content to place in which user's Timeline to provide the most interesting news to the users during their period of attention.

This matchmaking can be improved by understanding two determinants: the interests of users in content as well as by understanding the meaning of content. In the remainder of this Section, we first examine user's content contribution to understand Facebook's initial situation for placing content in timelines. We then explain the newsfeed arrangement which reflects Facebook's assumptions and wishes which content to view. The consumption of presented content is evaluated thereafter. Finally, we show the impact of the provided content based on the measured user reactions such as comments and likes. The following Timeline related evaluations are based on the subset of 774 users who viewed at least 100 posts (788938 in total) to avoid outliers to affect our results.

\subsection{Content Generation}
 \label{sec:content_generation}

The Timeline contains not only user generated content. Facebook itself is creating a big portion of posts that appear in the Timeline e.g. to inform users about status or profile picture updates of their friends. Companies (add pages) can also post regular newsfeed messages. However, in the remainder of this Section (\ref{sec:content_generation}), we focus on user-generated content rather than content generated from Facebook itself or commercial pages.

  \begin{figure}[ht]
\centering
\includegraphics[width=0.489\textwidth]{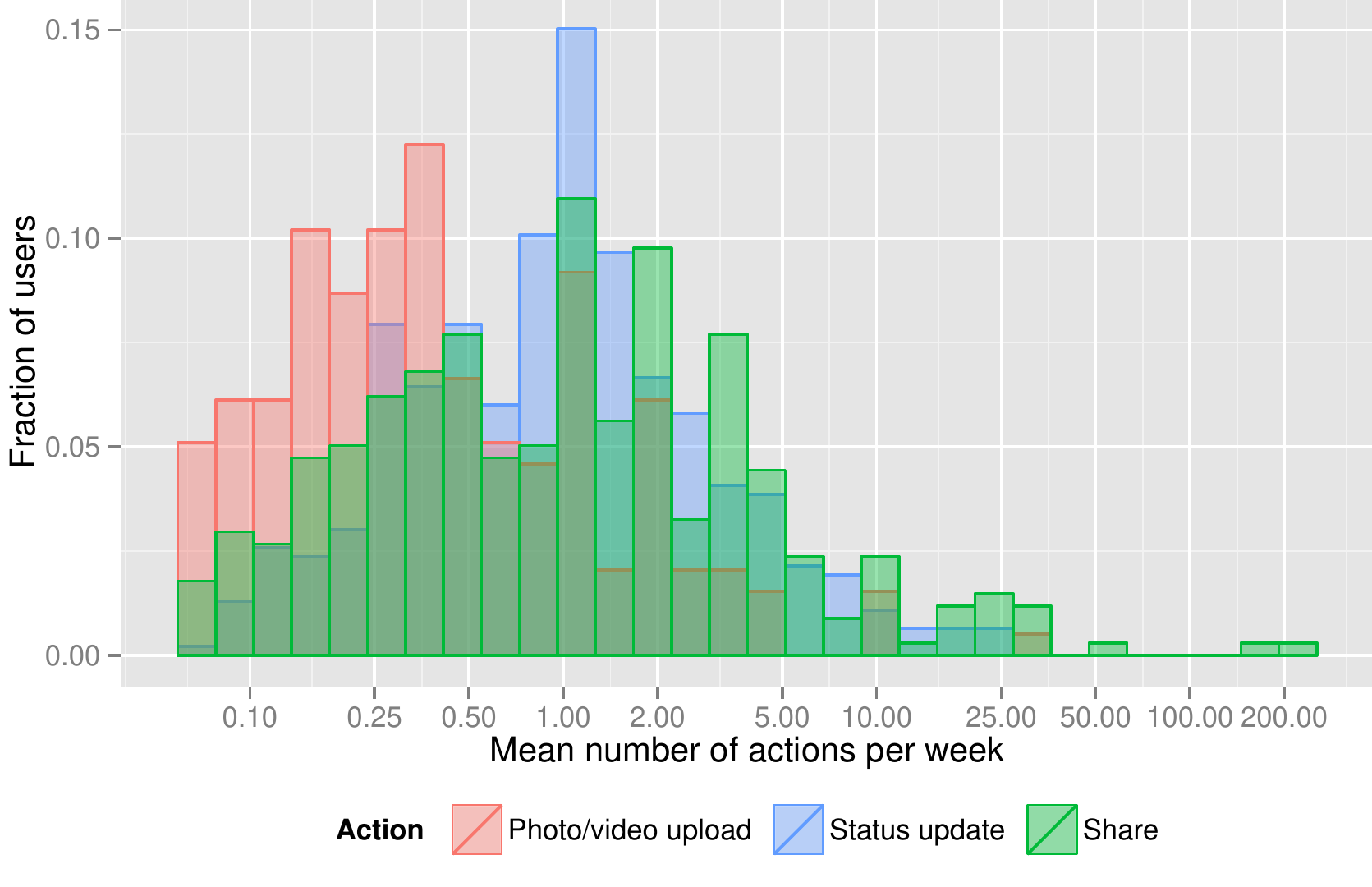}
\caption{Content type}
\label{fig:mean_number_posting_content_per_hour}
\end{figure} 

Users can create different types of content: status updates which only consist of pure text messages, shared links to internal or external pages as well as media such as photos or videos. In general, publishing content is not very popular. 36.37\% of the users who viewed at least 100 posts during the observation period did not post anything. We excluded them from further content generation evaluations.
 
 \begin{figure}[ht]
\centering
\includegraphics[width=0.48\textwidth]{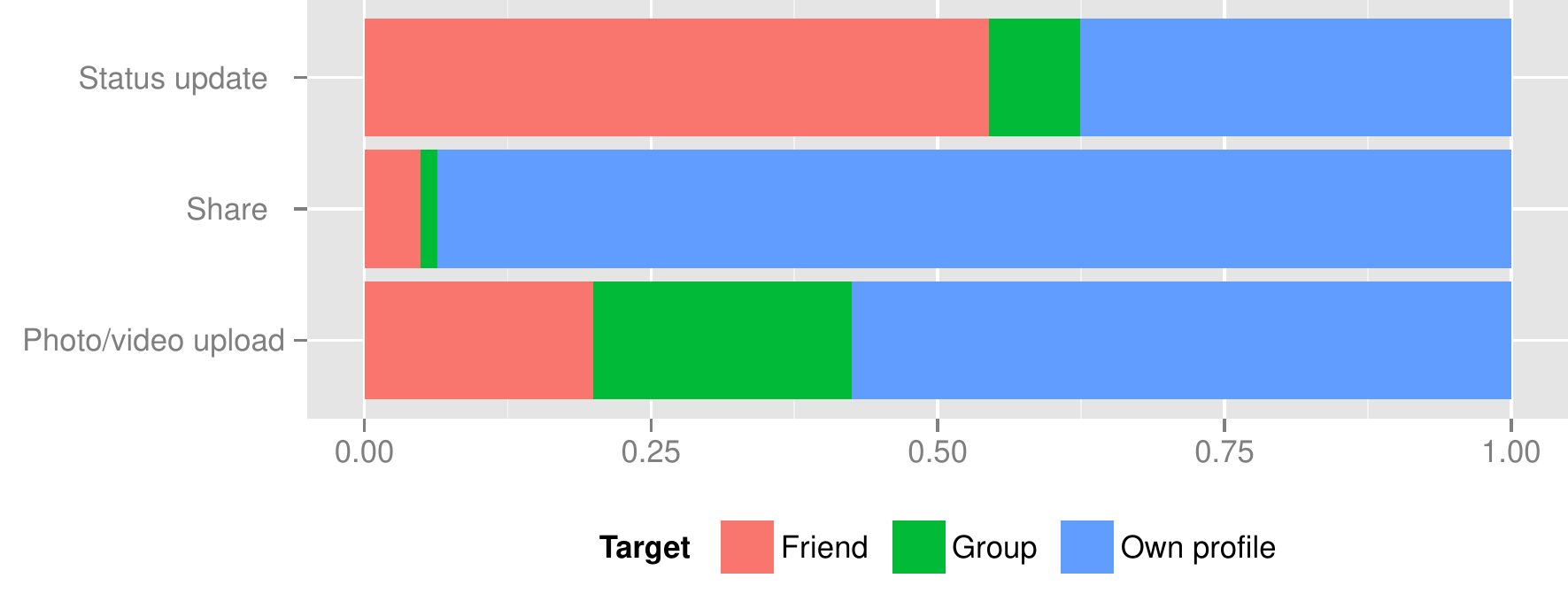}
\caption{Content posting targets}
\label{fig:target_of_posting_content}
\end{figure} 

Figure \ref{fig:mean_number_posting_content_per_hour} shows the popularity of the three kinds of content uploads. The main finding is that the average number of posts per week is extremely diverse with respect to different users. The relative frequency of sharing content can be represented by a long tailed distribution. For example, we observed users to share more than 200 links per week on average. 

However, the overwhelming majority only posts very few items. Our FPA users created on average 4.36 posts per week. Most frequently, they shared already existing content amongst others (2.43 times per week). Status updates and photo and video uploads happen less often. On average, 1.57 Status updates and 0.35 photos have been posted.

Beside the own profile, users can also leverage other channels to post content. Figure \ref{fig:target_of_posting_content} shows what type of content is posted on the own profile, at a friend's profile or on a group's newsfeed. The overwhelming majority of Links are  shared on the own Timeline and most photos and videos are published there as well. However, elaborating details regarding status update posts depict different patterns. Facebook invites users to send birthday congratulations via e-mail notification and provides a dedicated page for this purpose. The the result is that the majority of status updates has been published on friend's profiles whereof 36.42\% of them have been directly posted via the sidebar of birthday congratulation pages.

\subsection{Newsfeed Composition}
\label{sec:newsfeed}

In this Section, we evaluate the composition of the newsfeed. We examine who authors the newsfeed content to depict the nature of the service and we evaluate which fraction of friends contributes to the newsfeed to scrutinize Facebook's eligibility as a tool to keep in touch with friends. 

\begin{figure}[ht]
\centering
\includegraphics[width=0.45\textwidth]{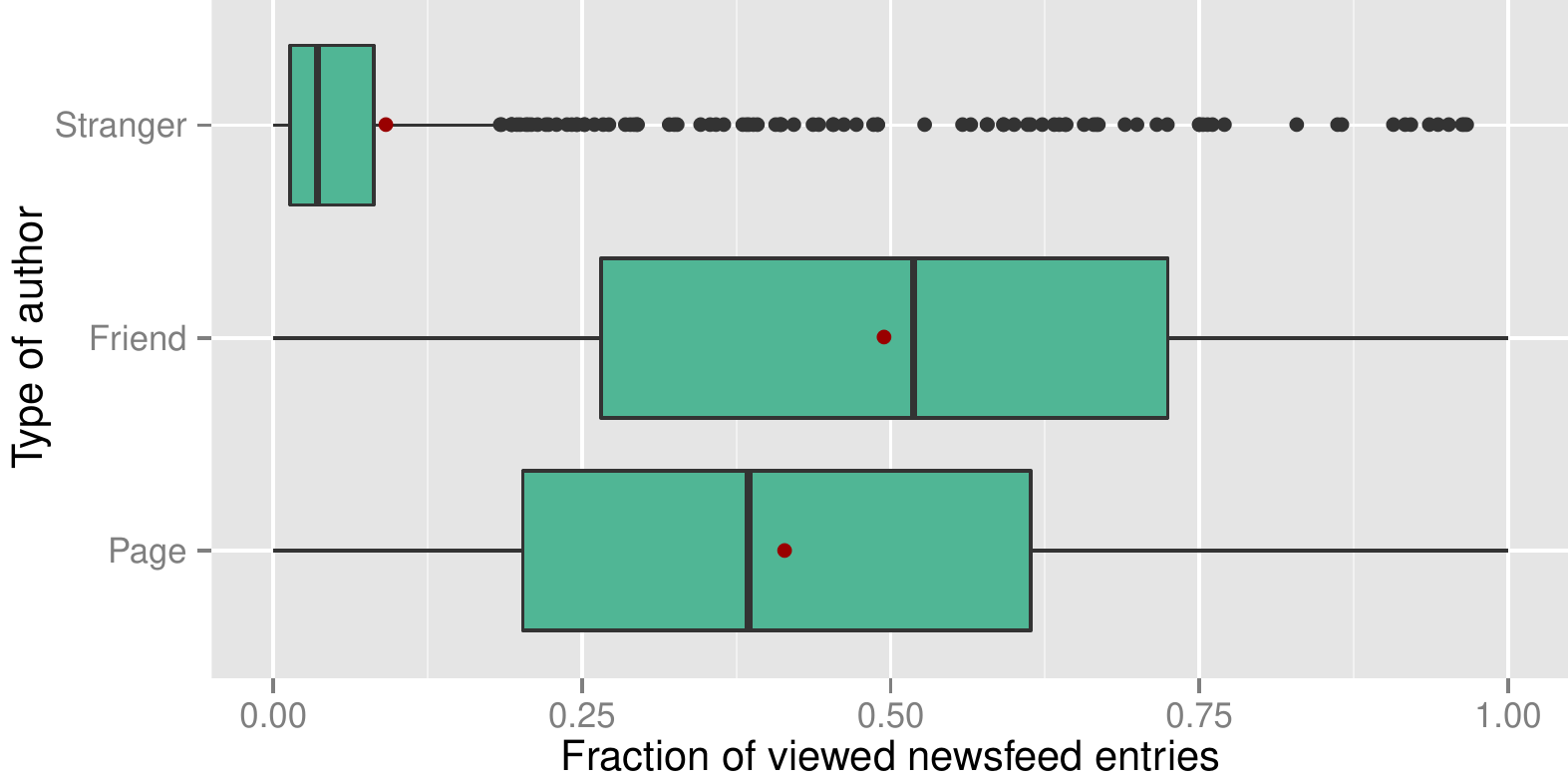}
\caption{Authors of newsfeed entries; red dots mark the averages}
\label{fig:authors}
\end{figure}

Facebook was initially planned for students to establish and maintain connections between friends. Figure \ref{fig:authors} shows the distribution of newsfeed entries with respect to the authorship. Nowadays, the average fraction of friend-generated content dropped to 49.5\%. While introducing professional pages as a tool for companies to communicate with their customers, Facebook became an important advertising platform that now reaches an average fraction of 41.4\% of commercial newsfeed entries. Only a very small fraction of content was initially posted by strangers (9.1\%). Content from strangers may appear in the personal newsfeed in case that friends interact with it (e.g. attach likes or comments).

The fraction of friends appearing in the newsfeed is critical for Facebook as a service that allows users to stay in touch with the whole set of friends. The intuitive assumption is that the fraction of friends that contributes to the newsfeed of users decreases with an increasing set of friends. This assumption is based on the facts that an increasing set of friends means a bigger set of stories for Facebook to choose for including into the Timeline while users only spend a limited amount of attention to the newsfeed. 

\begin{figure}[ht]
\centering
\includegraphics[width=0.489\textwidth]{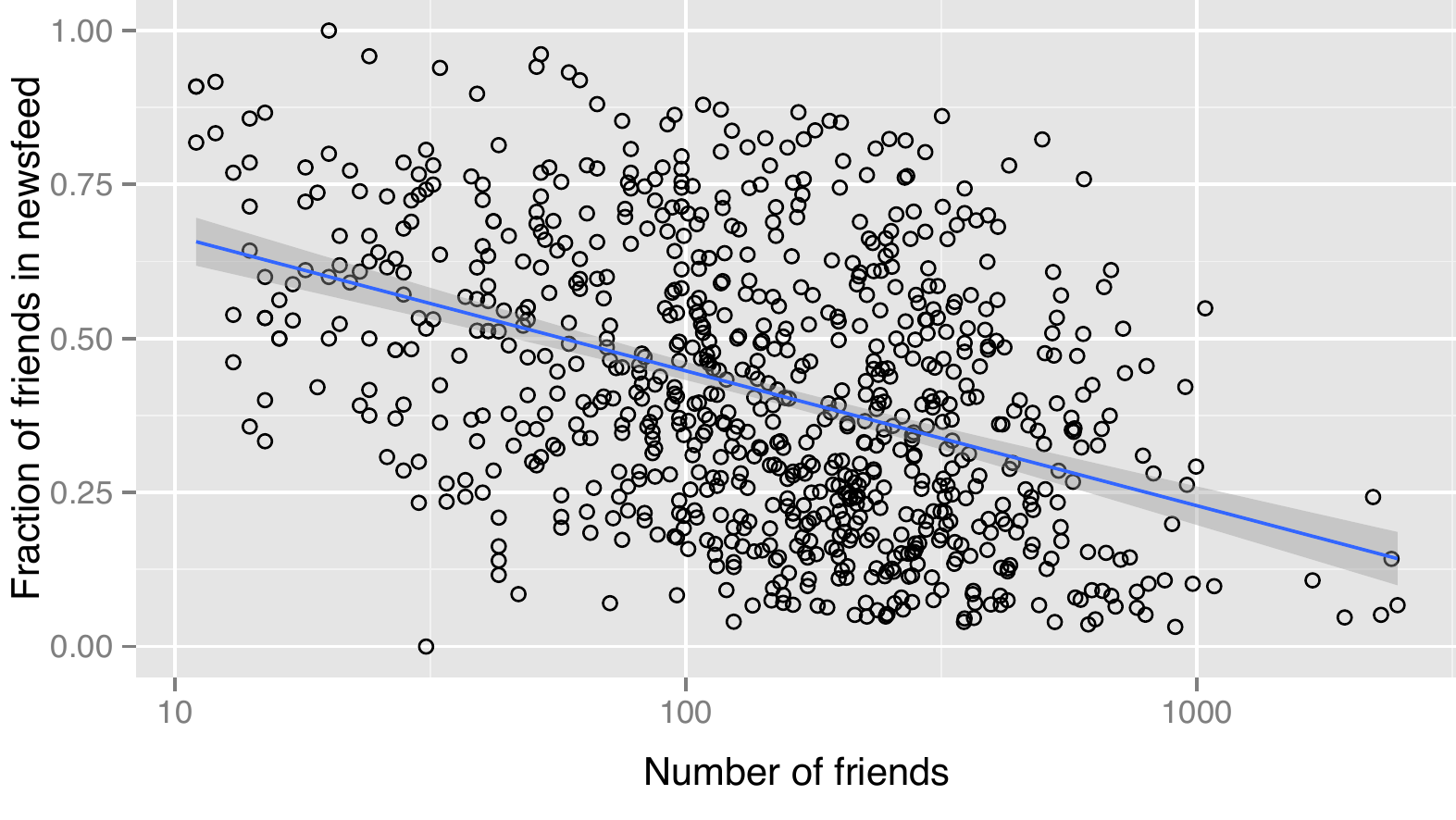}
\caption{Relation between the total number of friends and the fraction of friends appearing in the newsfeed (linear regression)}
\label{fig:regression_fraction_friends}
\end{figure}

Figure \ref{fig:regression_fraction_friends} shows the linear regression on the size of the set of friends versus the fraction of friends in the newsfeed. It shows both: that the decreasing assumption holds as well as that the relation is not very strong (-0.32). Our interpretation is that Facebook tries to include as many friends as possible into the Timeline. 

Some users with the huge set of more than 700 friends still see content of more than 50\% of their contacts in their newsfeed. This leads to a huge amount of newsfeed entries and suggests that Facebook scales the number of content items in the newsfeed according to the attention that it receives.

\begin{figure}[ht]
\centering
\includegraphics[width=0.489\textwidth]{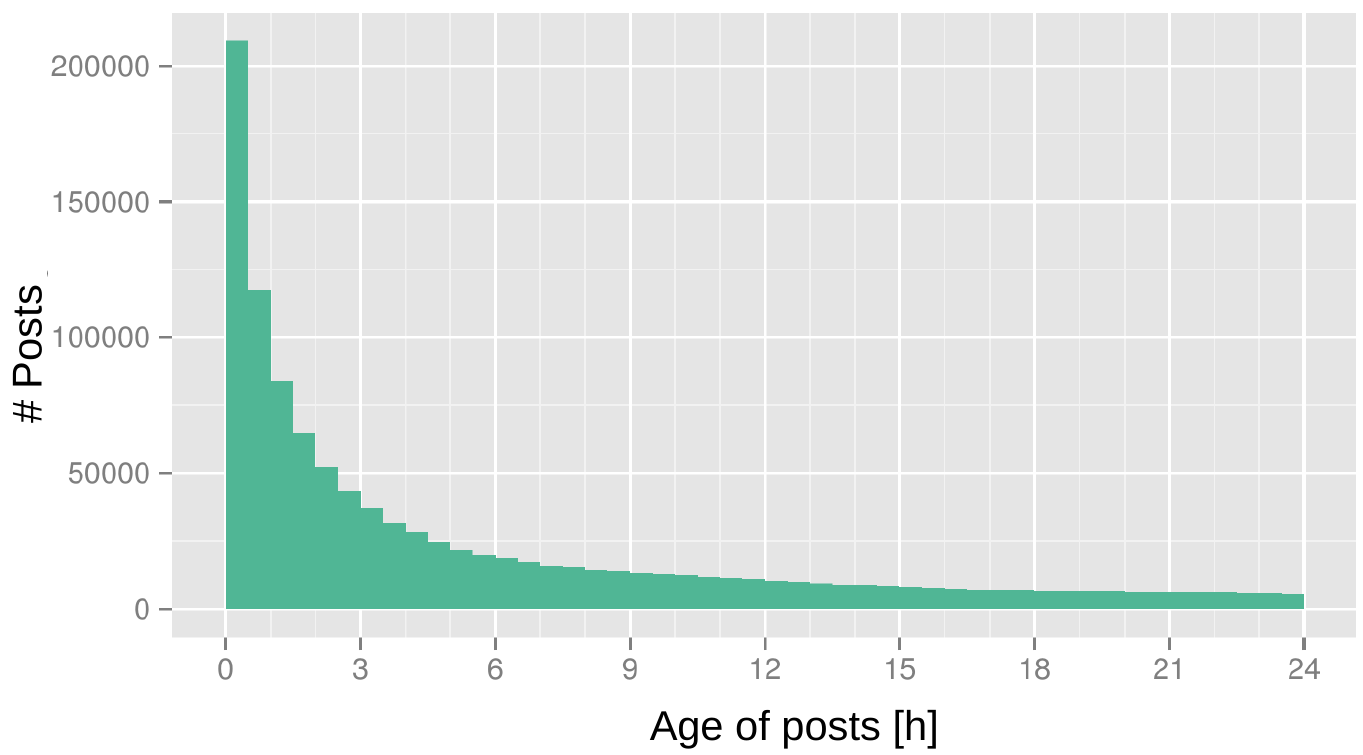}
\caption{Age of newsfeed posts until being displayed in hours}
\label{fig:age_newsfeed}
\end{figure} 

Most newsfeed entries are very fresh until they are displayed (Figure \ref{fig:age_newsfeed}): 84.79\% are not older than 24 hours and 25.77\% are created less than one hour before. However, a small fraction of entries is very old: 4.02\% are older than 7 days 1.73\% are older than 30 days. One reason for very old content to appear or reappear in the newsfeed are friends liking or commenting old content.

\subsection{Content Consumption}
\label{subsec:consumption}

In this Section, we provide insights into the content consumption habits of the FPA users. %
We first evaluate how long different types of newsfeed entries stay in the viewport of the browser before being clicked. This gives an idea about how much effort FPA users invest into the decision which content they view. We then evaluate how many newsfeed entries have been viewed on average per day to estimate the amount of attention a user pays to the newsfeed. Finally, we examine the types of accessed content to allow comparing the posted with the viewed content.

\begin{figure}[ht]
\centering
\includegraphics[width=0.489\textwidth]{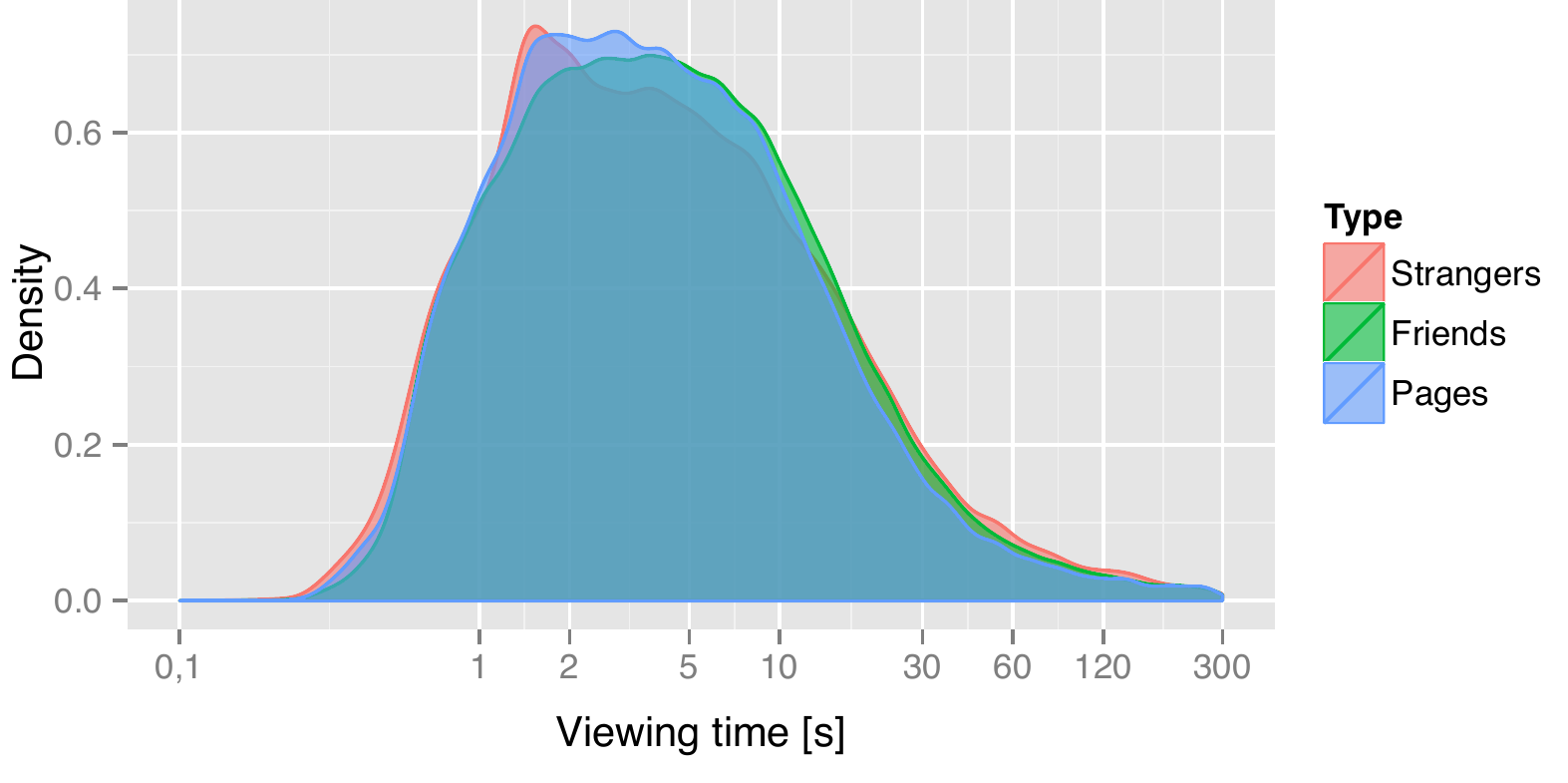}
\caption{The time that newsfeed entries stay in the browser viewport with respect to the authors of entries}
\label{fig:newsfeed_duration}
\end{figure}

Figure \ref{fig:newsfeed_duration} shows the time a newsfeed entry stays in the browser viewport before being clicked. This information illustrates the time investments of users to check a certain entry. Most users invest between one and ten seconds (average 9.5s) to decide whether to click on an entry or not. This is valid for all types of entries independent from the authorship. Very interesting is that posts from commercial pages receive a similar attention than posts from friends or strangers (9.8s vs. 9.1s). However, posts from strangers cause a slightly more diverse checking time than others. While there is a peak at 1.5 seconds, there are is also a higher value at 60 seconds compared with posts from friends or pages.

\begin{figure}[ht]
\centering
\includegraphics[width=0.489\textwidth]{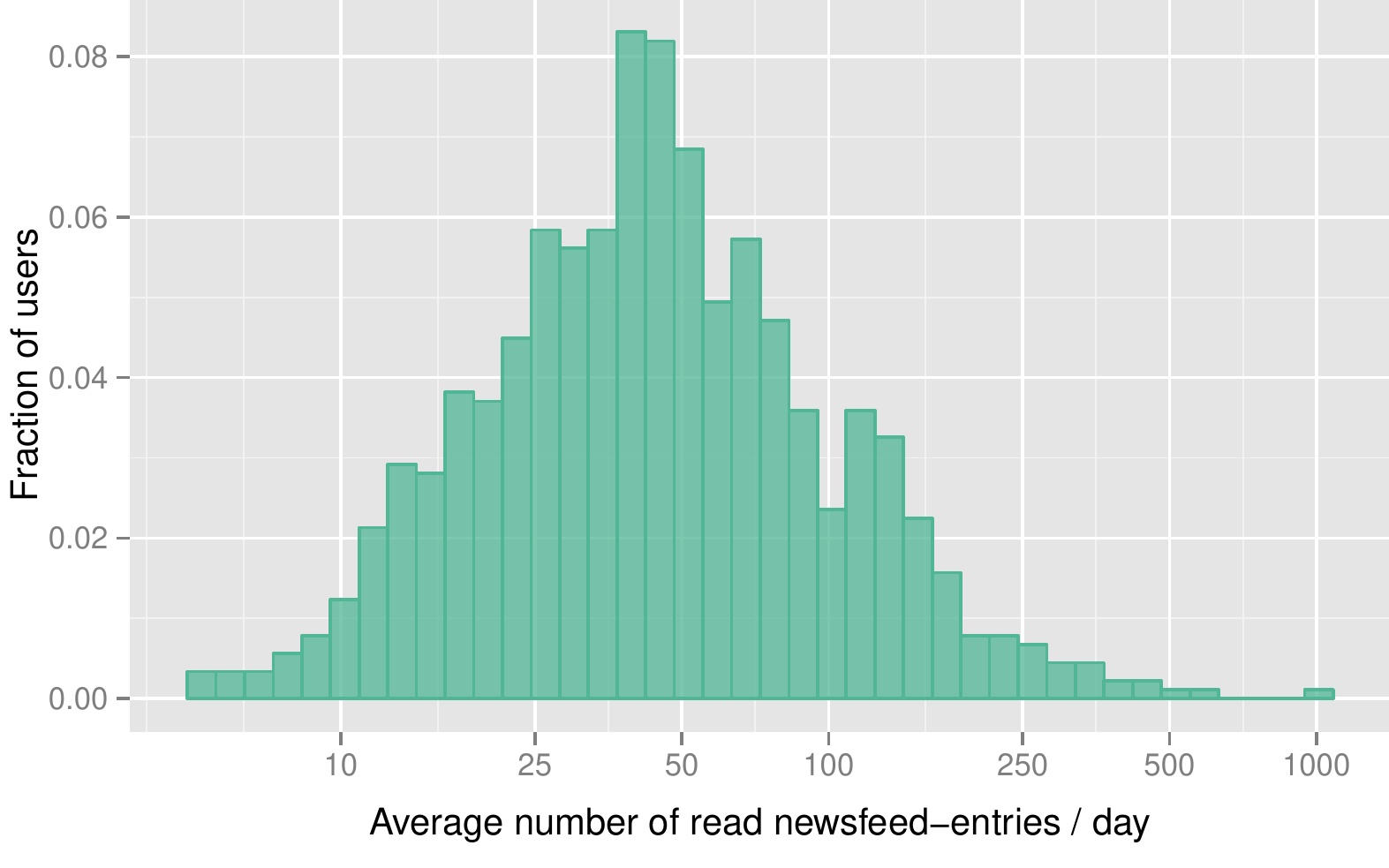}
\caption{Distribution (Histogram) of the average number of newsfeed entries that have been viewed by FPA users during the observation period; each bar indicates the fraction of users viewing the respective average number of posts}
\label{fig:number_entries}
\end{figure}

\begin{figure}[ht]
\centering
\includegraphics[width=0.489\textwidth]{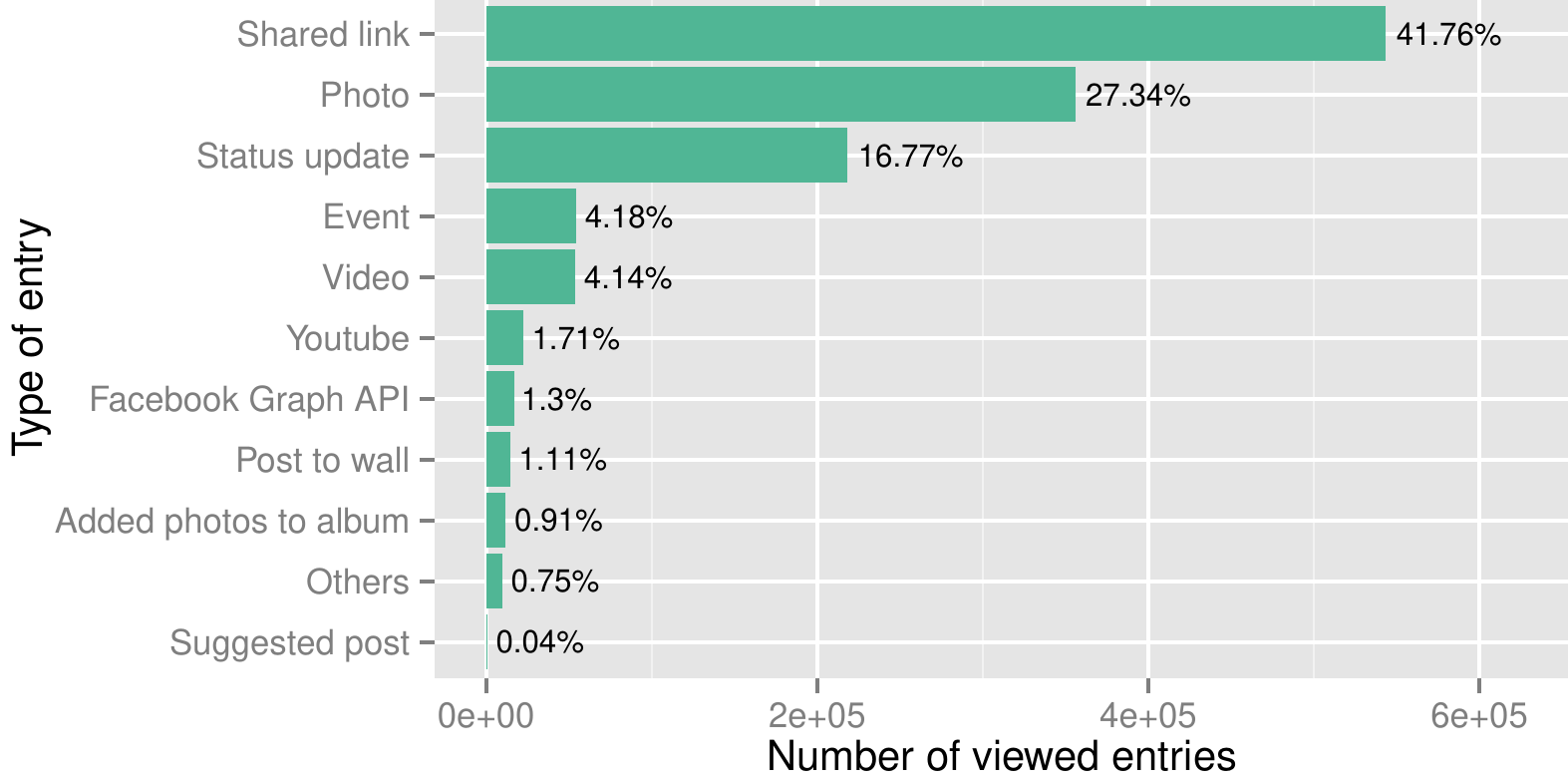}
\caption{Number and fraction of newsfeed entries that are viewed by FPA users}
\label{fig:newsfeed_entry_types}
\end{figure} 

On average, users view 43 posts per day. A histogram that allows to estimate the distribution can be found in Figure \ref{fig:number_entries}. Figure \ref{fig:newsfeed_entry_types} shows the composition of newsfeed entry views. The biggest fraction of viewed posts consists of shared links (41.76\%), photos (27.34\%) and status updates (16.77\%). Considering the authorship of posts, it is surprising for us that the clicked shares of commercial posts roughly equal those of friends or strangers. FPA users seem to accept commercial newsfeed posts equally as regular news beside user-generated content.

\begin{figure}[ht]
\centering
\includegraphics[width=0.489\textwidth]{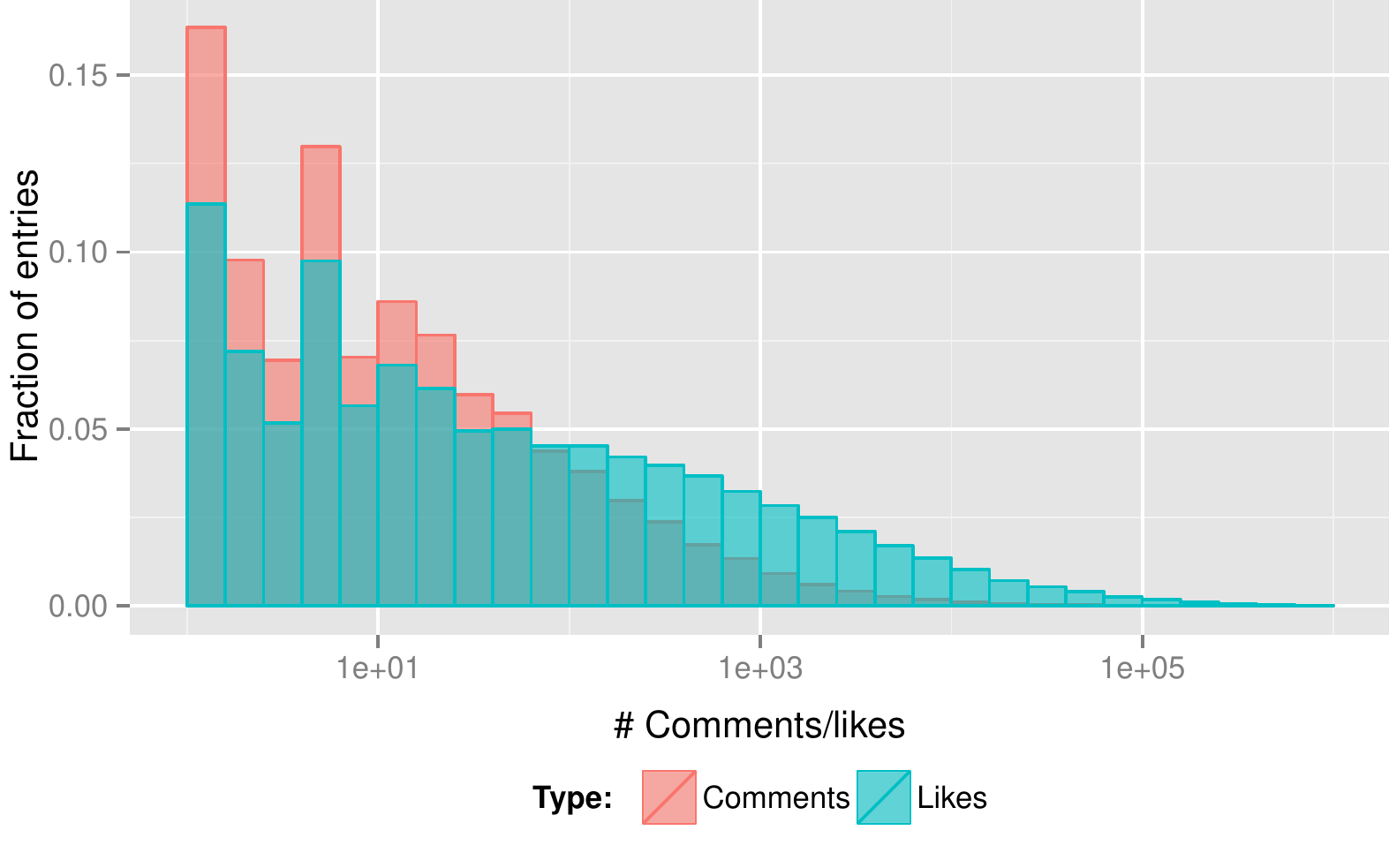}
\caption{Number of comments and likes of newsfeed entries}
\label{fig:comments_likes_number}
\end{figure}

Figure \ref{fig:comments_likes_number} shows the long tailed distributions of the number of comments, attached to newsfeed entries. This long tailed distribution shows the number of comments and likes to be extremely disparate amongst FPA users. On average, they like roughly 4\% and comment 1\% of all newsfeed posts.

\section{Communication Patterns} 
\label{sec:communication_patterns}

In this Section, we provide insights into communication patterns of FPA users. We first separately evaluate the user profile views as such representing the most popular two-sided communication functionality. In case of viewing user profiles, the profile owners are information sender and the user who is accessing the profile is the information receiver.

\subsection{User Profile Access}

\begin{figure}[ht]
\centering
\includegraphics[width=0.389\textwidth]{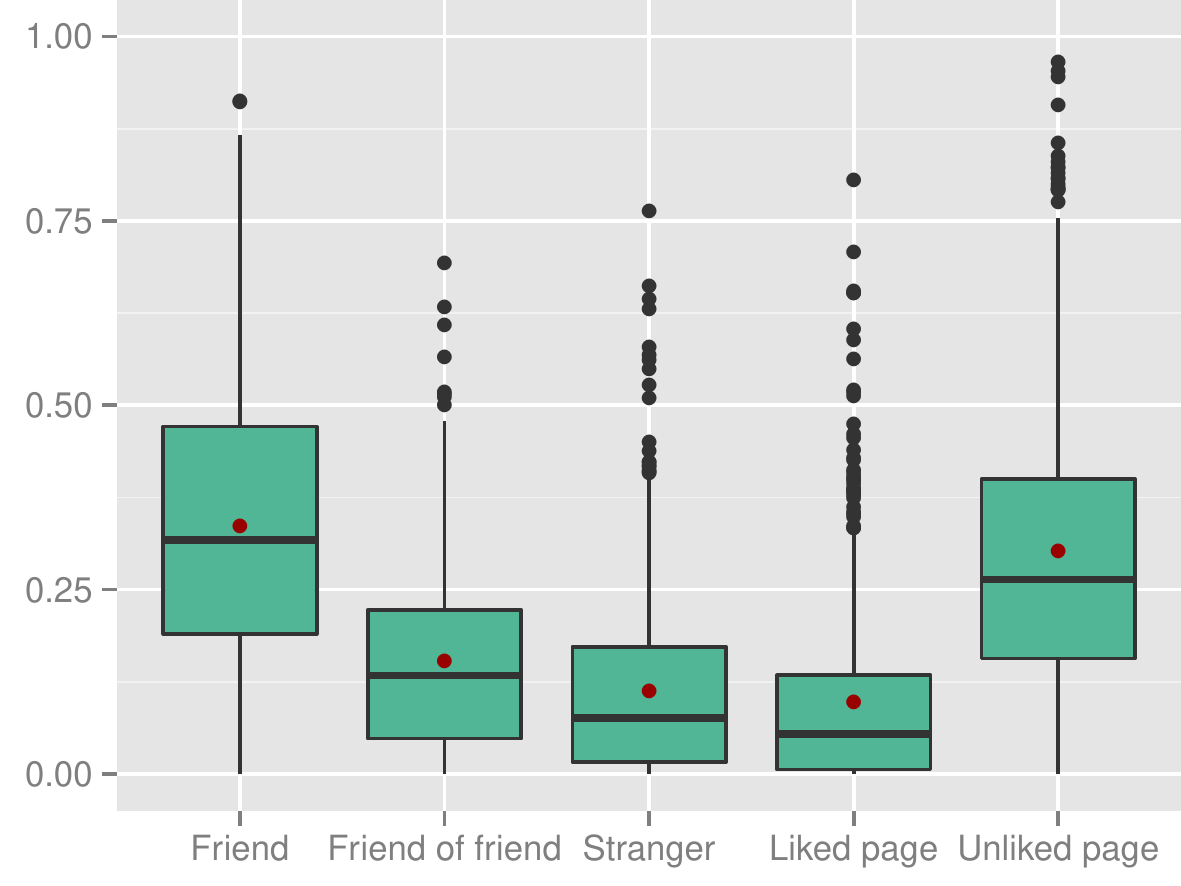}
\caption{Profile page access with respect to the social graph distance; we distinguish professional pages in liked and not liked (unliked) pages}
\label{fig:profile_access}
\end{figure} 

Profiles in Facebook can be classified into two major categories: profiles of individuals and profiles of professional pages that represent companies or prominent persons such as actors or musicians. In Figure \ref{fig:profile_access}, we further distinguish amongst friend's, friend-of-friend's and stranger's profiles as well as between liked and unliked professional pages. FPA users visit on average 33.51\% pages of friends, 30.19 \% 'unliked' professional pages 15.32\% pages of friend-of-friends, 11.22 \% stranger's pages and 9.76 \% liked professional pages.

\subsection{Communication with Friends}

\begin{figure}[ht]
\centering
\includegraphics[width=0.489\textwidth]{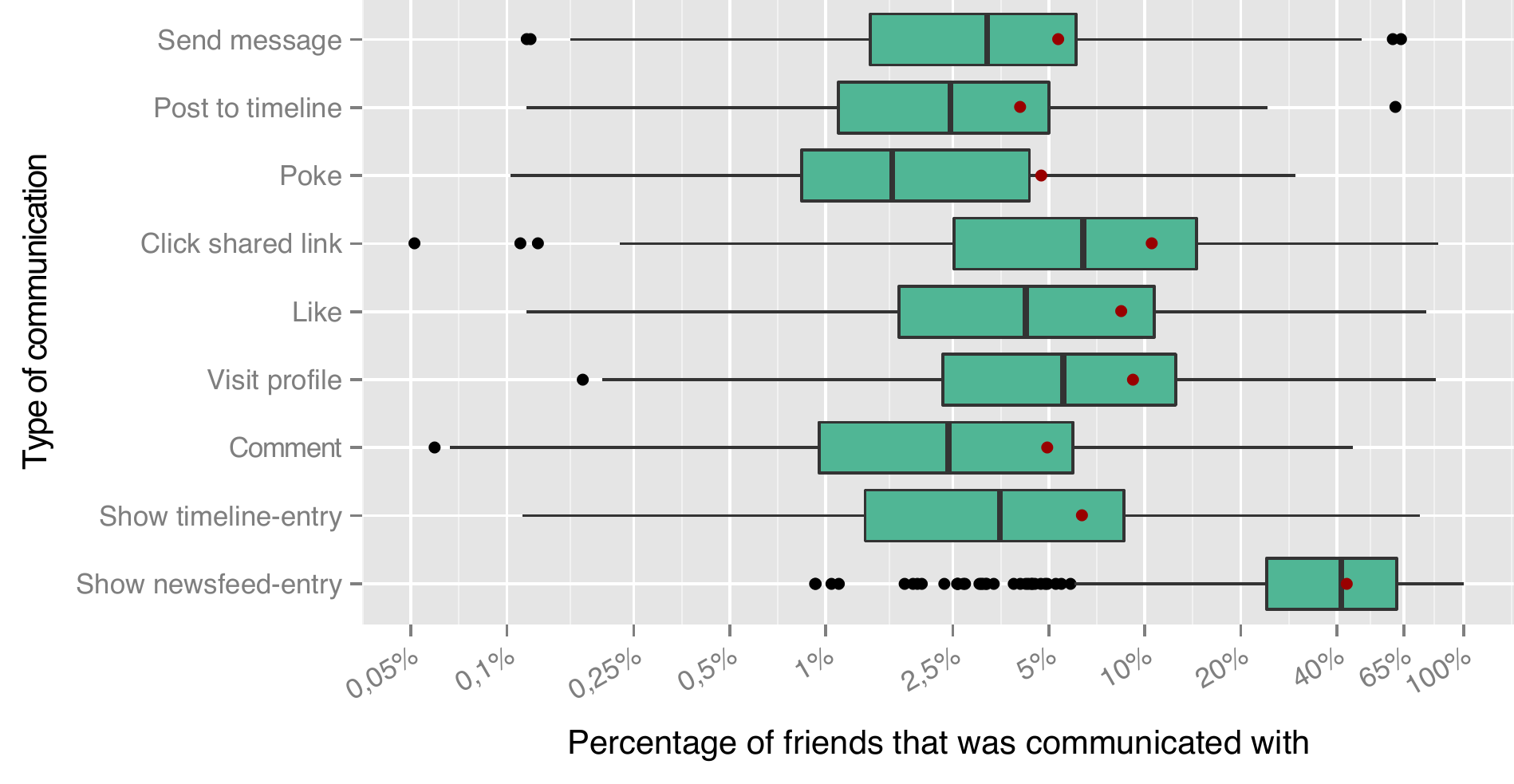}
\caption{Percentage of friends with whom the FPA users communicate with respect to the communication function; red dots mark averages}
\label{fig:commnication_with_percentage_of_friends}
\end{figure}

Facebook is widely known as a tool to communicate with friends. To understand this communication amongst friends, we elaborate the two-sided functionalities. Figure \ref{fig:commnication_with_percentage_of_friends} shows the percentages of friends that have been communicated with during the observation period by using a certain communication function. Since this analysis is affected by a too short observation time, we only included data from 714 users who participated for more than four weeks in our study. 

With outstanding advance, the most popular type of communication between friends is viewing newsfeed entries of each other, followed by clicking on shared links, viewing profiles and liking content. On the other side of the spectrum, least popular communication functions are poking, timeline posts and commenting. Excluding the extreme cases of poking and viewing newsfeed entries, the most popular functions are those which imply the lowest commitment of the acting user. 

Both poking and viewing newsfeed entries are exceptional cases for different reasons. Poking is exceptional because it is only intensively used by a very small fraction of users and the newsfeed is arranged by Facebook's algorithms without an explicit choice of the users. Facebook shows newsfeed entries of friends in case of communicating with them using any other function. %
The set of friends, appearing in the newsfeed, contains all other sets in Figure \ref{fig:commnication_with_percentage_of_friends}  almost completely (98\%). Thus the majority of FPA users communicates with only a minority of friends. This is especially true for users with more than 80 friends (Figure \ref{fig:regression_fraction_friends}: more dots are below the 50\% marker than above).

\section{Dynamics in User Behavior on Facebook} 
\label{sec:dynamics}

The social networking phenomena has recently appeared and the idea still disseminates amongst world's population. The establishment of best practices as well as the process of users to learn how to use social networking tools like Facebook is an ongoing process. Furthermore, Facebook is permanently working hard to improve the service. Thus, usage patterns evolve over time and examining user behavior in the field of social networking means to examine a quickly moving target. 

\begin{figure}[ht!]
\centering
\includegraphics[width=0.489\textwidth]{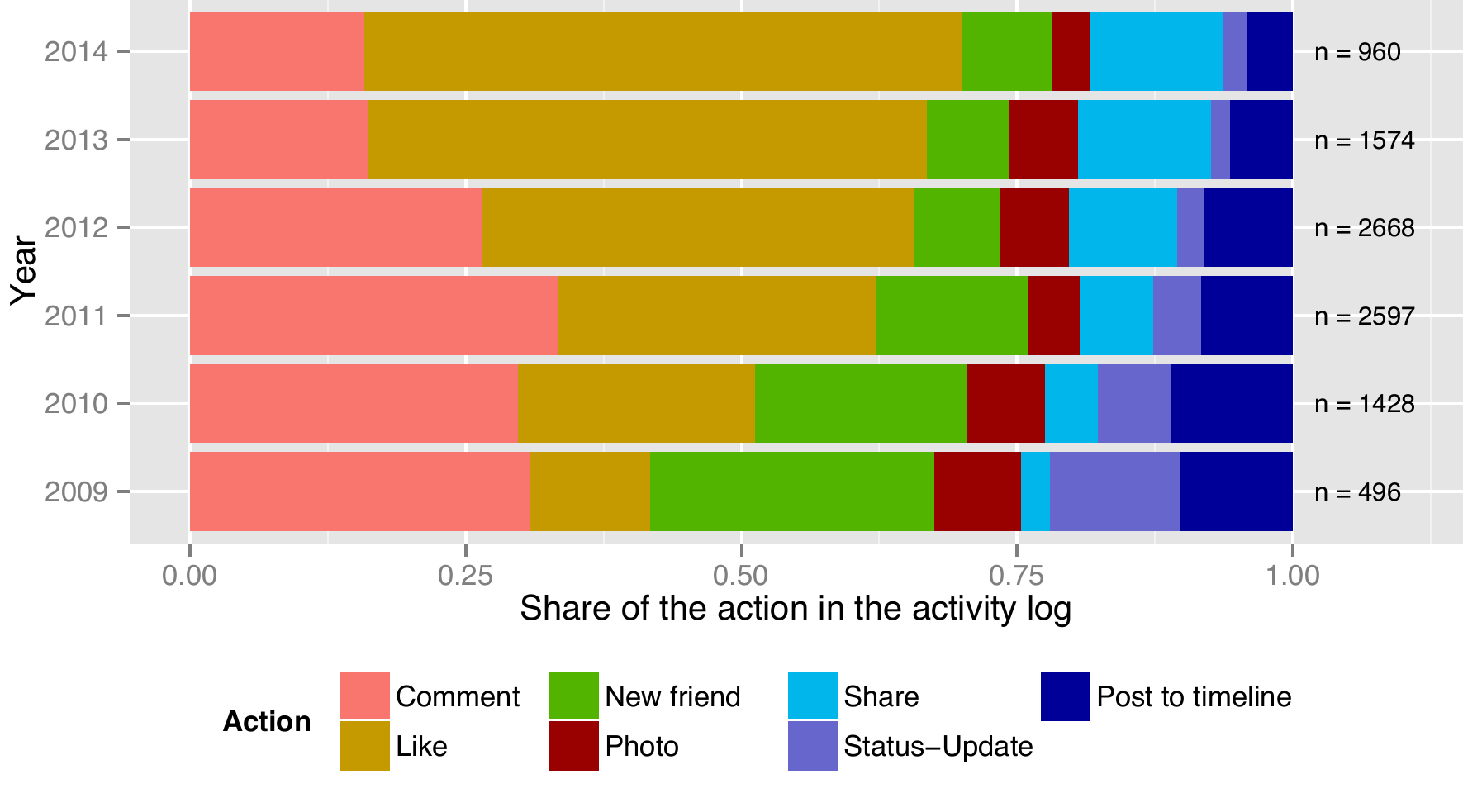}
\caption{Comparison of Facebook usage from 2009 till 2014}
\label{fig:actions_per_year}
\end{figure}

Fortunately for us, Facebook encloses an activity log into user profiles. It contains all actions of many categories that users have performed, starting from the day when a user is registering her account at Facebook. Our tool, FPA, is able to read this information. To show the development of usage patterns, we compare the popularity of the seven most popular activities on Facebook per year in Figure \ref{fig:actions_per_year}.

This plot strikingly showcases the maturity of Facebook and its decreased growth rates. The fraction of friend-adding actions dropped from year to year. This indicates that the process of new friends joining the network as well as the establishment of friendship connections converged to the natural social dynamics in the society.

Also noticeable is that the share of actions which require little effort from users increases: Likes and sharing of content recently became much more popular than in 2009. The fraction of sharing actions increased by factor 4.62 from 2009 till 2014. Accordingly, the fraction of comment, status update actions and photo uploads decreased.

\section{Summary and Conclusion}
\label{sec:conclusion}

In this paper, we presented a study on user behavior in Facebook, based on data from 2,071 participants. We elaborated the session durations, the average number of sessions per day and the function usage. The goal was to examine how intensive users use Facebook and to determine which is the dominating function of Facebook. We further evaluated the newsfeed and depicted what kind and how much information is shared amongst which actors. Since Facebook is widely known as a tool to communicate with friends, we checked this assumption by explicitly elaborating how often the study participants communicated with their friends using a certain functionality. Finally, we elaborated the user behavior from 2009 till 2014 to understand the recent development.

The evaluation of our dataset yielded several different findings. We summarize them in as follows:
\begin{itemize}
 \item Sessions in Facebook are shorter and less frequent than assumed in the literature. In particular, very long sessions are missing.
 \item The newsfeed is the most intensively used function of Facebook.
 \item Content contribution is very disparate. A few users contribute a major share of content.
 \item FPA users consume many items in a short time per day (average: 43 items in 6:44 minutes).
 \item Shared content in Facebook is very fresh. 84.79\% of all posts are not older than 24 hours until being shown to the recipients.
 \item %
 The probability of a commercial newsfeed entry to be viewed roughly equals those of friend's posts in average.%
 \item User behavior in Facebook is changing at the scale of years. While low effort actions, such as likes and reshares, recently became more popular, the contribution of photos, status updates and comments is decreasing. 
 \item Facebook became mature and stable. This is reflected not only by decreasing user growth rate but also by decreasing establishment of new connections amongst the existing set of users. Users discover fewer new people to add as friend.  
\end{itemize}

We conclude from these observations that users recently seem to use Facebook with a higher speed and lower effort than before, preferring quick actions with low commitment (e.g. likes and reshares). Also, users prefer extremely fresh content. As a consequence, alternative OSN architectures, such as P2P-based OSNs (e.g. \cite{cutillo09safebook-2,my3,s-data,buchegger09peerson-impl,paul2014survey}), could be designed in a lightweight way without the burden to persistently store stale content in large user profiles. Because of their low storage overhead, the large fraction of shared (external) links in the newsfeeds supports the idea of small user profiles, too. Furthermore, focus of alternative OSN architectures should be brought on dynamic environments, caused by short session durations.

This work has also highlighted the dynamics in user behavior at a scale of years. We assume both technological influences, such as advances in the sector of mobile computing, as well as the social reasons, e.g. learning curves of users and privacy discussions, to be drivers of dynamics in user behavior. User behavior in OSNs thus should be studied while being aware of dynamics and old user models should be carefully used to evaluate novel systems.

Surprisingly for us, users spend a major share of their attention and time with commercial pages. The probability of a commercial newsfeed entries to be viewed roughly equals those of friend's posts (does not hold for a small set of close friends). Thus, Facebook seems to be successful to target the recipients of commercial news and it seems to insert a compatible amount of commercial content into the newsfeed.

\section{Acknowledgements}
This work has been co-funded by the German Research Foundation (DFG) in the Collaborative Research Center (SFB) 1053 'MAKI – Multi-Mechanisms-Adaptation for the Future Internet.


\bibliographystyle{plain}
\bibliography{bib/Journal,bib/p2psn}

\end{document}